\documentclass[nofootinbib,preprintnumbers,amsmath,amssymb,twocolumn]{revtex4-2}

\usepackage{slashed}

\newcommand{\SUtet}{\boldsymbol{e}}
\newcommand{\suc}{\boldsymbol{S}} 

\begin{document}
	
\title{Reinterpreting Supersymmetry}
\author{Igor Salom}
\affiliation{Institute of Physics, Belgrade \\ University, Pregrevica 118, Zemun, Serbia }

\begin{abstract}
The mathematically elegant and promising idea of supersymmetry faces severe challenges as its conventional weak-scale realizations are increasingly constrained by the lack of experimental evidence of superpartners. However, we point out that the prediction of new (seemingly non-existing) supersymmetric particles is not a necessity: if we slightly relax our expectations from the supersymmetric models, generators $Q^a_\alpha$ of $N$-extended supersymmetry can be readily interpreted as carrying a quantum number of the internal gauge group, thus connecting particles with different gauge properties. In this view, operators $Q^a_\alpha$ remain ``square roots'' of translations, but do not by themselves generate symmetries of the model. They merely represent transformations that mathematically relate bosonic and fermionic fields; it is only the gauge-invariant combinations of the form $\sum_a \{Q^a_\alpha, \bar Q_{a\beta}\}$ that correspond to spacetime momenta and thus represent symmetries of the model. With this conceptual modification, the simplest Yang-Mills supersymmetric model no longer connects vector bosons with hypothetical gauginos, but with far less exotic chiral fermions, e.g.\ with left-handed leptons. Further adding an $SU(2)$ doublet of Higgs scalars no longer introduces Higgsinos, but a fermion that naturally corresponds to the right-handed lepton, with the familiar Yukawa term showing up as a mathematical necessity. Despite giving up the requirement that operators $Q^a_\alpha$ alone generate symmetries of the action, this approach to supersymmetry is strikingly mathematically similar to standard SUSY, raising hopes that many of the favorable properties of standard supersymmetry can be retained, while potentially reconciling the idea of supersymmetry with experimental data. The approach is still quite restrictive: the relative coefficients of such models are strongly determined.
\end{abstract}

\maketitle

\section{Introduction}

Supersymmetry posits the existence of transformations that map bosonic fields into fermionic and vice versa. At the same time these transformations should leave the action invariant (and, under the standard assumptions, correspond to symmetries of a relativistic $S$-matrix), as it is natural for any proper ``symmetry''. This last requirement invokes Haag-Lopuszanski-Sohnius theorem \cite{HLS} and highly constrains the possible form of SUSY algebra. For $N$-extended Poincare supersymmetry in four dimensions, this set of requirements dictates standard superalgebra relations 
\cite{SohniusSUSY, WessBaggerSUSYbook}:   
\begin{eqnarray}
	&& \bigl\{ Q^a_R, 
	\overline{Q_R}_b\bigr\}
	=2\delta^a_b
	\bigl(P_R \gamma^\mu \bigr) P_\mu , \label{anticommQQP}\\
	&& \bigl\{Q_{R}^a, Q_{R}^b\bigr\} = 2\bigl(P_R C^{-1}\bigr) Z^{ab}, \label{anticommZ}\\
	&& \bigl[P_\mu, Q^a \bigr] = 0, \label{commPQ}
\end{eqnarray}
with central charges $Z^{ab}= -Z^{ba}$, $P_{R,L} = (1 \pm \gamma_5)/2$, 
$Q_R = P_R Q$, and $\overline{Q_R} = (Q_R)^\dagger \gamma^0
 = \bar Q P_L$.
 
Despite initial historical enthusiasm \cite{SohniusSUSY}, nowadays it is conventionally accepted that it is not possible to identify the R-symmetry (that transforms different supercharges into each other) with any internal symmetry gauge group \cite{Rsymmetry} (though gauged R-symmetries do occur, notably in supergravity, and some other attempts persist, e.g.\ \cite{RAttemptExample}). Instead, the gauge groups of internal symmetries are taken to commute with supersymmetry generators, and that consequently leads to predictions of new particles, in apparent conflict with experimental results \cite{PDG:2024,SUSYproblems:2025}.

In an attempt to resolve this problem, we here propose a radical step: to discard requirements (\ref{anticommZ}) and (\ref{commPQ}) -- and along with them the premise that individual $Q$ operators generate symmetries of the theory, while retaining only the essential concept that supersymmetries square to the translations, i.e.\ (\ref{anticommQQP}). This allows the following modification of the remaining relation: 
\begin{equation}
	\left\{ Q^a_R, 
	\overline{Q_R}_a \right\}
	=2\bigl(P_R \gamma^\mu \bigr) P_\mu , \label{anticommQQPnew}
\end{equation}
where instead of $\delta^a_b$ we now have the extended supersymmetry index contracted. The goal of this paper is to demonstrate that after relaxing supersymmetry requirements to merely (\ref{anticommQQPnew}) it becomes natural to
identify the internal symmetry gauge group with the R-symmetry (or its subgroup), i.e.\ to interpret indices of the extended-supersymmetry as the gauge group indices. And to show that, surprisingly, much of the mathematical structure and at least some of the valuable features of supersymmetric theories remain, while the conflict with experimental data is potentially avoided.

In this setting, that we may call {\it relaxed supersymmetry}, 
only the gauge invariant (trace) combinations of anticommutators 
(fully contracted over the internal labels) are constrained: by the defining relation (\ref{anticommQQPnew}) such combinations must equal a combination of momenta, and, as such, they represent symmetries of the model. All other anticommutators of supersymmetry generators remain {\it a priori} unconstrained by relaxed supersymmetry. In particular, the absence of Kronecker delta from (\ref{anticommQQPnew}) implies that operators of the form $\{Q^a_R, \overline{Q_R}_b \}$, with $a \neq b$ are not required to be zero, and the same is true for anticommutators $\bigl\{Q_R^a, Q_R^b\bigr\}$ that also here remain unspecified and potentially non-central. The existence of such bosonic operators would have been in conflict with the Coleman-Mandula theorem, had we retained (\ref{commPQ}) and insisted that operators $Q$ represent symmetries in the standard sense. 

The paper is organized as follows. In the next section we will show that in relaxed supersymmetry Yang-Mills vector fields (belonging to the adjoint representation of the gauge group) and chiral leptons (belonging to the fundamental representation) naturally belong to the same supermultiplet. (We retain the terminology such as ``supercharge'' for $Q_R^a$ and ``supermultiplet'' for collections of fields they relate, although individually these operators are neither conserved nor generators of symmetries of the action.) We will see that the mathematical form of the relations is nevertheless very similar to that of the standard supersymmetry, but with the benefit that the construction no longer requires the introduction of experimentally unobserved gauginos. In addition, we will show that the total numbers of bosonic and fermionic degrees of freedom must still match, and this will require introduction of precisely $N$ families of fermions, where $N$ matches the rank of the unitary gauge group.

In the third section we will additionally introduce gauge invariant chiral leptons, and show that relaxed supersymmetry maps them onto scalar fields belonging to the fundamental representation. Combining both multiplets together (one containing Yang-Mills field and the one with scalar fields) necessarily yields familiar Yukawa terms, that force interpretation of the gauge invariant fermions as leptons of opposite chirality (i.e.\ if Yang-Mills field was paired to left-handed leptons, the scalar field is mapped to its right-handed counterparts). In $N=2$ case, the result will thus closely resemble the lepton-Higgs sector of the Standard Model. 

While the goal of the sections II and III will be to illustrate -- as simply and as directly as possible -- the main ideas and implications of the relaxed supersymmetry, Section IV deals with some hitherto ignored subtleties related to the gauge transformations of the supersymmetric generators $Q^a_\alpha$. More specifically, whereas the supersymmetric generators in sections II and III will be effectively treated in a fixed gauge, in Section IV we will discuss a more gauge covariant formulation.

Section V summarizes the results.

\section{Yang-Mills multiplet}

In order to emphasize both similarities and differences between the standard and relaxed supersymmetry we will briefly recapitulate some of the well known and here relevant SUSY relations, written in a suitable notation. Since the similarity is actually exhibited between the relaxed supersymmetry with $N$ generators and non-extended standard supersymmetric models with $U(N)$ gauge group, we will start from the usual non-extended supersymmetric action for a $U(N)$ Yang-Mills vector field $A_\mu$ and a Majorana gaugino field $\Psi$ \cite{Ferrara:SYM}: 
\begin{equation}
	\begin{aligned}
		S_{\mathrm{YM}}^s
		=
		\int d^4x&\,
		\Bigl[
		-\frac12
		(F_{\mu\nu})^a{}_b
		(F^{\mu\nu})^b{}_a\\
		&+
		i\,
		(\overline{\Psi_L})^a{}_b
		\gamma^\mu
		(D_\mu \Psi_L)^b{}_a
		\Bigr]. \label{SinSYM}
	\end{aligned}
\end{equation}
To facilitate later comparison, we expressed the action using gauge group indices $a$ and $b$ in the fundamental representation and conveniently normalized left-handed projection of the gaugino field $\Psi_L\equiv \sqrt2\,P_L\Psi$. The superscript $s$ indicates ``standard supersymmetry''. The corresponding canonical Hamiltonian can be written as a sum of the gauge invariant part and the Gauss part \cite{Dirac:1964:HamiltonianAnalysis,Henneaux:HamiltonianAnalysis}:
\begin{equation}
	H_{\mathrm C}^s = H_{\mathrm{inv}}^s + H_{\mathrm G}^s,
\end{equation}
where
\begin{equation}
	\begin{aligned}
		H_{\mathrm{inv}}^s
		=
		&
		\int d^3x\,
		\Big[
		(E_i)^a{}_b(E_i)^b{}_a
		+
		(B_{i})^a{}_b(B_{i})^b{}_a
		\\
		&
		\qquad -i(\overline{\Psi_L})^a{}_b
		\gamma^i(D_i\Psi_L)^b{}_a
		\Big],\\
		H_{\mathrm G}^s = &-2\int d^3x\, (A_0)^a{}_b(\mathcal G)^b{}_a,
	\end{aligned} \label{HinSYM}
\end{equation}
$E_i=F_{0i}$, $B_i=\tfrac12\varepsilon_{ijk}F_{jk}$ (with
$\varepsilon_{123}=+1$), $\eta_{\mu\nu}=\mathrm{diag}(+,-,-,-)$ and the first-class Gauss constraint is $\mathcal G^a{}_b = (\mathcal G_{\mathrm{YM}}^s)^a{}_b \equiv (D_i E_i)^a{}_b -\frac g2
\big[
(\Psi_L^\dagger)^a{}_c
(\Psi_L)^c{}_{b}
-
(\Psi_L^\dagger)^c{}_b
(\Psi_L)^a{}_{c}
\big] \simeq 0$.
 
Analogously, for the spatial momenta:
\begin{eqnarray}
	&& P_k^{s, \mathrm C} = 	P_k^{s, \mathrm{inv}}+P_k^{s, \mathrm G},\nonumber\\
	&& P_k^{s, \mathrm{inv}} = \int d^3x\,
		\left[
		\varepsilon_{ki\ell} \{
		(E_i)^a{}_b,(B_\ell)^b{}_a\}
		+ 	i(\Psi^\dagger_L)^a{}_b
		(D_k\Psi_L)^b{}_a
		\right],\nonumber\\
	&& P_k^{s, \mathrm G}
	=
	-2\int d^3x\,
	(A_k)^a{}_b(\mathcal G)^b{}_a. \label{PinSYM}
\end{eqnarray}

After imposing usual equal-time commutation relations
\begin{eqnarray}
	&&\left[
	(A_i)^a{}_b(t,\mathbf x),
	(E_j)^c{}_d(t,\mathbf y)
	\right]
	=
	\frac{i}{2}
	\delta_{ij}
	\delta^a{}_d\delta^c{}_b
	\delta^3(\mathbf x-\mathbf y),\nonumber\\
	&&
	\left\{
	(\Psi_L)_\alpha{}^a{}_b(t,\mathbf x),
	(\Psi_L^\dagger)^\beta{}^c{}_d(t,\mathbf y)
	\right\}
	=
	(P_L)_{\alpha}{}^\beta
	\delta^a{}_d\delta^c{}_b
	\delta^3(\mathbf x-\mathbf y),\nonumber
\end{eqnarray}
supercharges given by:
\begin{equation}
	Q_R
	=
	\frac1{\sqrt 2}\int d^3x\, 
	(F_{\mu\nu})^a{}_b \,
	i \gamma^{\mu\nu}\gamma^0\Psi_L^b{}_a,
	\label{QSYM}
\end{equation}
with $\gamma^{\mu\nu}\equiv\tfrac12[\gamma^\mu,\gamma^\nu]$, satisfy non-extended case of relation (\ref{anticommQQP}) weakly, i.e.\ up to the terms proportional to the first class constraint $\mathcal G$:
\begin{equation}
	\left\{
	Q_R,\overline{Q_R}
	\right\}
	=
	2(P_R \gamma^\mu)
	P_\mu^{s,\mathrm{inv}}  
	\simeq 
	2(P_R \gamma^\mu)
	P_\mu^{s,\mathrm{C}}. \label{QQPinv}
\end{equation}
(Or, as often stated, the anticommutator of two supersymmetric transformations yields a translation plus a gauge transformation.) Here and below, operator identities are, as usual, understood modulo total spatial derivatives (fields assumed to fall off sufficiently fast at spatial infinity). The charges generate the following standard transformations of the fields:
\begin{equation}
	\begin{aligned}
		\left[
		(Q_R)_\alpha,
		(A_i)^a{}_b
		\right]
		&=
		\frac{1}{\sqrt2}
		(\gamma_i \Psi_L^a{}_b)_\alpha,
		\\[4pt]
		\left\{
		(\overline{Q_R})^\alpha,
		(\Psi_L)^a{}_{b\beta}
		\right\}
		&=
		\frac1{\sqrt 2}
		(F_{\mu\nu})^a{}_b
		(P_L i\gamma^{\mu\nu})_\beta^{\ \alpha}.
	\end{aligned}
	\label{transformationsSYM}
\end{equation}
In addition, the charges $Q_\alpha$ are both conserved, i.e.\ $[H, Q_\alpha]=0$, and translation invariant $[P_k, Q_\alpha]=0$, reflecting that the theory is symmetric with respect to above field transformations. 

The essential observation in this paper is that almost identical structure also exists if we consider a Yang-Mills field interacting with a chiral spinor in fundamental representation $(\Psi_L)_a$  (instead of the Majorana gaugino in the adjoint $(\Psi_L)^a{}_b$):
\begin{equation}
	S_{\mathrm{YM}}
	=
	\int d^4x\,
	\left[
	-\frac12
	(F_{\mu\nu})^a{}_b
	(F^{\mu\nu})^b{}_a
	+
	i\,
	(\overline{\Psi_L})_a
	\gamma^\mu
	(D_\mu\Psi_L)^a
	\right].\label{SinYM}
\end{equation}
The absence of the other index on spinor fields is the only difference in comparison with expression (\ref{SinSYM}). Consequently, Hamiltonian and momenta also have almost identical forms as in the supersymmetric case. Their gauge invariant parts are:
\begin{equation}
	\begin{aligned}
		H_{\mathrm{inv}}
		=
		&
		\int d^3x\,
		\Big[
		(E_i)^a{}_b(E_i)^b{}_a
		+
		(B_{i})^a{}_b(B_{i})^b{}_a
		\\
		&
		\qquad -i(\bar\Psi_L)_a
		\gamma^i(D_i\Psi_L)^a
		\Big],\\
		P_k^{\mathrm{inv}} = &
		\int d^3x\,
		\left[
		\varepsilon_{ki\ell}
		\{(E_i)^a{}_b,(B_\ell)^b{}_a\}
		+ 	i(\Psi_L^\dagger)_a
		(D_k\Psi_L)^a
		\right]. \label{HPinYM}
	\end{aligned}
\end{equation}
The generic forms of $H_{\mathrm G}$ and $P_k^{\mathrm G}$ remain the same as in (\ref{HinSYM}-\ref{PinSYM}), with an expected slight modification of the Gauss law: 
$(\mathcal G_{\mathrm{YM}})^a{}_b = (D_i E_i)^a{}_b + \frac g2 (\Psi_L^\dagger)_{b} (\Psi_L)^a \simeq 0$.

If we now modify generators (\ref{QSYM}) so that they also carry a gauge group index (by simply removing one index from the fermion field), and normalize them accordingly: 
\begin{equation}
	Q_R^a
	= \frac 1{\sqrt{2 N}}
	\int d^3x\,
	(F_{\mu\nu})^a{}_b \,
	i \gamma^{\mu\nu}\gamma^0\Psi_L^b ,
\end{equation}
they will satisfy:
\begin{eqnarray}
	&& \frac 14 (\gamma_0)_\beta{}^\alpha \left\{ Q^a_\alpha, 
	\overline{Q}_{a}^\beta\right\} =
	\int d^3x\,
	\Big[\frac 1N \Big(
	(E_i)^a{}_b(E_i)^b{}_a
	\nonumber \\
	&&
	\qquad +
	(B_{i})^a{}_b(B_{i})^b{}_a\Big) -i (\bar\Psi_L)_a
	\gamma^i(\widehat D_i\Psi_L)^a
	\Big],\\
	&& \frac 14 (\gamma_k)_\beta{}^\alpha \left\{ Q^a_\alpha, 
	\overline{Q}_{a}^\beta\right\} = 
	\int d^3x\,
	\Big[
	\frac 1N \varepsilon_{ki\ell}
	\{(E_i)^a{}_b,(B_\ell)^b{}_a\} \nonumber\\
	&& 
	\qquad + i (\Psi_L^\dagger)_a
	(\widehat D_k\Psi_L)^a
	\Big],
\end{eqnarray}
where $(\widehat D_i\Psi_L)^a$ is (traceless) $SU(N)$ part of the covariant derivative, i.e.:
\begin{equation}
	(\widehat D_i\Psi_L)^a = (\delta^a_b \partial_i - i g \widehat A_i^a{}_b) \Psi_L^b; 
	\quad \widehat A_i^a{}_b = A_i^a{}_b - \frac 1N \delta_b^a A_i^c{}_c.
\end{equation}

By comparing with the expressions for gauge invariant momenta (\ref{HPinYM}), we see two differences preventing (\ref{anticommQQPnew}) to be (weakly) satisfied: the traceless covariant derivative of fermion fields and the $1/N$ factors appearing in the bosonic energy and momentum densities. None of them is introduced by hand: the traceless connection $\widehat D_i$  emerges from the contraction of gauge indices in the
anticommutator, the trace part of the connection cancelling between the
two contractions; the factor $1/N$ multiplies only the bosonic
densities, because in the fermionic term the closed gauge-index
contraction $\delta^a{}_a=N$ arising from the commutator of the two field
strengths supplies a compensating factor $N$.

The first difference requires only a minor adjustment: in this simplest model, the algebra thus requires fields $\Psi_L$ to carry zero $U(1)$ charge. Importantly, notice also that this is no different from the supersymmetric case (\ref{SinSYM}-\ref{PinSYM}): $U(1)$ part of the connection is not interacting with gaugino, so we might have used $\widehat D_\mu \Psi_L$ instead of $D_\mu \Psi_L$  already in these expressions. However, the $1/N$ multiplier requires a more significant modification, since it cannot be removed by simply rescaling the fields or the generators. Its existence effectively reflects the imbalance between the numbers of fermionic and bosonic degrees of freedom \cite{Dreiner:freedomCounting}. The most straightforward way to remove this mismatch is to introduce $N_{\mathrm{g}}$ copies of fermion fields -- i.e.\ $N_{\mathrm{g}}$ generations. It turns out that a specific choice $N_{\mathrm{g}} = N$ will solve this problem. Namely, taking both modifications into account, the expressions (\ref{SinYM}) and (\ref{HPinYM}) become:
\begin{equation}
	S_{\mathrm{YM}}^r
	=
	\int d^4x\,
	\left[
	-\frac12
	(F_{\mu\nu})^a{}_b
	(F^{\mu\nu})^b{}_a
	+
	i\,
	(\bar\Psi_L)_{aA}
	\gamma^\mu
	(\widehat D_\mu\Psi_L)^{aA}
	\right],\label{SinYMN}
\end{equation}
\begin{equation}
	\begin{aligned}
		H_{\mathrm{inv}}^r
		=
		&
		\int d^3x\,
		\Big[
		(E_i)^a{}_b(E_i)^b{}_a
		+
		(B_{i})^a{}_b(B_{i})^b{}_a
		\\
		&
		\qquad -i(\bar\Psi_L)_{aA}
		\gamma^i(\widehat D_i\Psi_L)^{aA}
		\Big],\\
		P_k^{r, \mathrm{inv}} = &
		\int d^3x\,
		\left[
		\varepsilon_{ki\ell}
		\{(E_i)^a{}_b,(B_\ell)^b{}_a\}
		+ 	i(\Psi_L^\dagger)_{aA}
		(\widehat D_k\Psi_L)^{aA}
		\right], \label{HPinYMN}
	\end{aligned}
\end{equation}
where summation is now implied also over the generation index $A = 1, 2, \dots N_{\mathrm{g}}$. Also $(\mathcal G_{\mathrm{YM}}^r)^a{}_b = (D_i E_i)^a{}_b + \frac g2 \left((\Psi_L^\dagger)_{b A} (\Psi_L)^{a A} - \frac 1N \delta^a_b (\Psi_L^\dagger)_{c A} (\Psi_L)^{c A} \right) \simeq 0$. Superscript $r$ denotes expressions exhibiting relaxed supersymmetry. Now, provided that we choose $N_{\mathrm{g}} = N$, operators defined as:
\begin{equation}
	Q_R^{aA}
	= \frac 1{\sqrt{2 N}}
	\int d^3x\,
	(F_{\mu\nu})^a{}_b
	\,
	i \gamma^{\mu\nu}\gamma^0\Psi_L^{bA}
	\label{QinYMN}
\end{equation}
finally satisfy: 
\begin{equation}
	\left\{ Q^{aA}_R, 
	\overline{Q_R}_{aA} \right\}
	=2\bigl(P_R \gamma^\mu \bigr) P^{r,\mathrm{inv}}_\mu 
	\simeq 2\bigl(P_R \gamma^\mu \bigr) P^{r,\mathrm{C}}_\mu . \label{QQPinvnew}
\end{equation}
That is: the anticommutator closes on the gauge-invariant four-momentum $P^{r,\mathrm{inv}}_\mu$, which is weakly equivalent to the canonical momentum $P^{r,\mathrm{C}}_\mu$. Thus the defining relation (\ref{anticommQQPnew}), with added summation over the generation index, is satisfied up to a gauge transformation, exactly as in standard supersymmetric Yang--Mills theory (\ref{QQPinv}). The factor $1/N$ in the bosonic densities was this time compensated by the sum over fermion species, $\sum_A\delta^A_A = N_{\mathrm{g}} = N$. (In principle, instead of labelling generations, $A$ index can be related to a subgroup of a larger gauge group, whose representation is indexed by the pair $aA$.) 

Thus, we can again express four-momenta as quadratic functions of $Q_R^{aA}$ (up to the Gauss constraint, as in the standard supersymmetry). Therefore, while not being supersymmetric in the standard sense, we see that the action (\ref{SinYMN}), involving Yang-Mills vector fields with $N$ generations of chiral fermions in fundamental representation, is supersymmetric in this new, relaxed sense.

Generators $Q_\alpha^{aA}$ induce transformations quite similar to those of the standard non-extended super-Yang-Mills case (\ref{transformationsSYM}):
\begin{equation}
	\begin{aligned}
		\left[
		(Q_R^{aA})_\alpha,
		(A_i)^b{}_c
		\right]
		&=
		\frac{\delta^a_c}{\sqrt{2N}}(\gamma_i \Psi_L^{bA})_\alpha,  \\
		\left\{
		(\overline{Q_R})^\alpha_{aA},
		(\Psi_L^{bB})_{\beta}
		\right\}
		&=
		\frac{\delta_A^B}{\sqrt{2N}}
		(F_{\mu\nu})^b{}_a
		(P_L i\gamma^{\mu\nu})_\beta^{\ \alpha}. 
	\end{aligned}
	\label{transformationsRSYM}
\end{equation}
The essential difference is that fermions are now in the fundamental representation, just as in the Standard Model. However, we stress that these transformations do not leave the action (\ref{SinYMN}) invariant. This is reflected in the fact that charges (\ref{QinYMN}) are not conserved, i.e.\ calculation shows that $[H^r, Q^{aA}_\alpha] \neq 0$ (although they are nonetheless translation-invariant by construction). Importantly, notice that in the abelian $U(1)$ case, both the standard (non-extended) supersymmetric $U(N)$ Yang-Mills model and its ``relaxed'' counterpart reduce to the same abelian model, supersymmetric in the standard sense. As we will discuss in more detail below, this allows us to see a certain subclass of relaxed supersymmetry simply as another way to generalize the (non-extended) supersymmetric Yang-Mills from the abelian to non-abelian case.

We should also note that, while generators $Q_\alpha^{aA}$, as defined by (\ref{QinYMN}), manifestly transform covariantly under the global gauge transformations, their transformation properties with respect to local transformations are a more complex issue, that we postpone until section IV. At this point we just note that the sum (\ref{QQPinvnew}) is fully gauge invariant, despite non-elegant gauge properties of $Q_\alpha^{aA}$ operators alone.

\section{Higgs multiplet}

The next step is introduction of a chiral (i.e. scalar-fermion) supermultiplet, in addition to the Yang-Mills (vector) multiplet. In standard (non-extended) SUSY, this is the way to obtain super Yang-Mills theory with matter \cite{deWit:1975}. There, it comes at the cost of introducing both gauginos and sfermions, neither of which seems compatible with present experimental data. The basic form of such supersymmetric action -- containing an anti-chiral multiplet transforming in the antifundamental $U(N)$ representation -- written in our notation becomes:
\begin{equation}
	\begin{aligned}
		S_{\mathrm{YMC}}^s &= \int d^4x\,
		\Big\{
		-\frac12
		(F_{\mu\nu})^a{}_b
		(F^{\mu\nu})^b{}_a
		+
		(D_\mu\phi^\dagger)^a
		(D^\mu\phi)_a
		\\
		&+
		i(\overline{\Psi_L})^a{}_b
		\gamma^\mu
		(\widehat D_\mu\Psi_L)^b{}_a
		+
		i(\overline{\psi_R})^a
		\gamma^\mu(D_\mu\psi_R)_a
		\\
		&-
		g\left[
		\phi^{\dagger a}
		(\overline{\Psi_L})^b{}_a
		(\psi_R)_b
		+
		(\overline{\psi_R})^a
		(\Psi_L)^b{}_a
		\phi_b
		\right]
		\\
		&-
		\frac{g^2}{4}
		\left(\phi^{\dagger a}\phi_a\right)^2
		\Big\}.
	\end{aligned}
	\label{SinSYMC}
\end{equation}
(This time we made it manifest that $U(1)$ part does not interact with gauginos.) The first four of the terms are kinetic -- one for each of the fields. It is the closure of algebra (\ref{anticommQQP}) -- even without the ``symmetry demands'' (\ref{commPQ}) -- that already requires introduction of the remaining terms. Thus, as we will shortly see, the relaxation of supersymmetry will only allow for a more general form of the terms, while retaining the overall rigid structure and the number of required terms.   
The corresponding invariant parts of the Hamiltonian and momenta are: 
\begin{equation}
	\begin{aligned}
		H_{\rm inv}^s
		=
		\int d^3x\,\Big\{&
		(E_i)^a{}_b(E_i)^b{}_a
		+
		(B_i)^a{}_b(B_i)^b{}_a
		\\
		&+
		\pi^\dagger_a \pi^a
		+
		(D_i\phi^\dagger)^a(D_i\phi)_a
		\\
		&-
		i(\overline{\Psi_L})^a{}_b
		\gamma^i(\widehat D_i\Psi_L)^b{}_a
		-
		i(\overline{\psi_R})^a
		\gamma^i(D_i\psi_R)_a
		\\
		&+ 
		g\left[
		(\overline{\Psi_L})^a{}_b
		(\psi_R)_a
		\phi^{\dagger b}
		+
		\phi_a
		(\overline{\psi_R})^b
		(\Psi_L)^a{}_b
		\right]
		\\
		&+
		\frac{g^2}{4}
		(\phi^{\dagger a}\phi_a)^2
		\Big\}, \\
		P_k^{s,\mathrm{inv}}
		=
		\int d^3x\,\Big[
		&
		\varepsilon_{ki\ell}
		\bigl(\{E_i,B_\ell\}\bigr)^a{}_a
		+
		i(\Psi_L^\dagger)^a{}_b{}
		(\widehat D_k\Psi_L)^b{}_{a}
		\\
		+
		\pi^a(D_k\phi)_a
		&+
		(D_k\phi^\dagger)^a\pi^\dagger_a
		+
		i(\psi_R^\dagger)^a
		(D_k\psi_R)_a{}
		\Big], \label{HPinSYMmat}
	\end{aligned}
\end{equation}
while Gauss constraint has additional contributions from new fields:
\begin{equation}
	(\mathcal G_{\mathrm{YMC}}^s)^a{}_b
	= (\mathcal G_{\mathrm{YM}}^s)^a{}_b -
	\frac g2
	(\psi^{\dagger}_R)^a
	(\psi_R)_b
	-
	\frac{ig}{2}
	\left[
	\phi^{\dagger a}\pi^\dagger_b
	-
	\pi^a\phi_b
	\right].
\end{equation}

Supersymmetric charges are:
\begin{equation}
	\begin{aligned}
		&Q_R
		=
		\frac{1}{\sqrt2}
		\int d^3x\,\Big\{
		(F_{\mu\nu})^a{}_b
		i \gamma^{\mu\nu}\gamma^0
		\Psi_L^b{}_a
		\\
		&-
		2 i
		(D_\mu\phi^\dagger)^a
		\gamma^\mu\gamma^0 \psi_R{}_a
		-
		g \phi^{\dagger a}\phi_b
		\gamma^0\Psi_L^b{}_a
		\Big\},
	\end{aligned}
	\label{QinSYMmat}
\end{equation}
with $(D_0 \phi^\dagger)^a	= \pi^a$. While the first two terms give the four kinetic terms in the Hamiltonian (\ref{HPinSYMmat}), the last term is required to cancel unwanted terms appearing in the $Q-Q$ anticommutators, in order to satisfy (\ref{anticommQQP}). 

And again we can obtain similar overall structure by considering 
relaxed-supersymmetry generators that nontrivially transform under gauge group and connect fermions with bosons of different gauge properties. We construct the relaxed supersymmetry generators from (\ref{QinSYMmat}) essentially by simply removing the lower index from both fermion fields -- turning gauginos into chiral fermions in the fundamental representation, and right-handed fermions in anti-fundamental representation into $SU(N)$-invariant fields: 
\begin{equation}
	\begin{aligned}
		Q_R^{aA}
		&=
		\frac{1}{\sqrt{2N}}
		\int d^3x\,\Big\{
		(F_{\mu\nu})^a{}_b
		i \gamma^{\mu\nu}\gamma^0
		\Psi_L^{bA}{}
		\\
		&-
		2 i 
		(D_\mu\phi^\dagger)^a
		\gamma^\mu\gamma^0 \psi_R^A
		-
		g \phi^{\dagger a}\phi_b
		\gamma^0\Psi_L^{bA}{}
		\Big\}.
	\end{aligned}
	\label{QinRSYMC}
\end{equation}
In addition, we again had to introduce $N_{\mathrm{g}}=N$ generations of fermions, to compensate for the difference in the number of fermionic and bosonic degrees of freedom. 

A direct calculation gives that these operators satisfy (\ref{QQPinvnew}), with: 
\begin{equation}
	\begin{aligned}
		H_{\rm inv}^r
		&=
		\int d^3x\,\Big\{
		(E_i)^a{}_b(E_i)^b{}_a
		+
		(B_i)^a{}_b(B_i)^b{}_a
		\\
		&+
		\pi^\dagger_a \pi^a
		+
		(D_i\phi^\dagger)^a(D_i\phi)_a
		\\
		&-
		i(\overline{\Psi_L})_{aA}
		\gamma^i(\widehat D_i\Psi_L)^{aA}
		-
		i(\overline{\psi_R})_A
		\gamma^i(D_i\psi_R)^A
		\\
		&+ 
		g\left[
		\phi^{\dagger a}
		(\overline{\Psi_L})_{aA}
		(\psi_R)^A
		+
		(\overline{\psi_R})_A
		(\Psi_L)^{a A}
		\phi_a
		\right]
		\\
		&+
		\frac{g^2}{4}
		(\phi^{\dagger a}\phi_a)^2
		\Big\}, \\
		P_k^{r,\mathrm{inv}}
		&=
		\int d^3x\,\Big[
		\varepsilon_{ki\ell}
		\bigl(\{E_i,B_\ell\}\bigr)^a{}_a
		+
		i(\Psi_L^\dagger)_{aA}
		(\widehat D_k\Psi_L)^{aA}
		\\
		&+
		\pi^a(D_k\phi)_a
		+
		(D_k\phi^\dagger)^a\pi^\dagger_a
		+
		i(\psi_R^\dagger)_A
		(D_k\psi_R)^A
		\Big]. \label{HPinRYMmat}
	\end{aligned}
\end{equation}
With  
$(\mathcal G_{\mathrm{YMC}}^r)^a{}_b
= (\mathcal G_{\mathrm{YM}}^r)^a{}_b - \frac{g}{2N}\,\delta^a_b\,
(\psi_R^\dagger)_A(\psi_R)^A - \frac{ig}{2} \big[
\phi^{\dagger a}\pi^\dagger_b - \pi^a\phi_b \big]
\;\simeq\; 0$
this corresponds to the following action:
\begin{equation}
	\begin{aligned}
		S_{\mathrm{YMC}}^r &= \int d^4x\,
		\Big\{
		-\frac12
		(F_{\mu\nu})^a{}_b
		(F^{\mu\nu})^b{}_a
		+
		(D_\mu\phi^\dagger)^a
		(D^\mu\phi)_a
		\\
		&+
		i(\overline{\Psi_L})_{aA}
		\gamma^\mu
		(\widehat D_\mu\Psi_L)^{aA}
		+
		i(\overline{\psi_R})_A
		\gamma^\mu(D_\mu\psi_R)^A
		\\
		&-
		g\left[
		\phi^{\dagger a}
		(\overline{\Psi_L})_{aA}
		(\psi_R)^A
		+
		(\overline{\psi_R})_A
		(\Psi_L)^{aA}
		\phi_a
		\right]
		\\
		&-
		\frac{g^2}{4}
		\left(\phi^{\dagger a}\phi_a\right)^2
		\Big\}. 
	\end{aligned}\label{SinRSYMC}
\end{equation}

And this we recognize as a quite common form of action, similar to what we have in the Standard Model, with no exotic matter. Only $N$ generations of fermion fields (that become massive, provided a suitable vacuum expectation value for $\phi$ is generated), interacting both with a $U(N)$ Yang-Mills field and with a Higgs-like field (via usual Yukawa interaction terms). The $\phi^4$ self-interaction term is also present. Such a model possesses the relaxed supersymmetry, while being free of potentially unphysical gaugino and sfermion fields required for standard supersymmetry. In $N=2$ case, the field content and interaction pattern closely resembles the lepton-Higgs sector of the minimal Standard Model \cite{Peskin:1995:HiggsLeptonSector} (i.e.\ without right-handed neutrinos), with precisely the left $SU(2)$ lepton doublet $\Psi_L$, right-handed singlet $\psi_R$, and a Higgs doublet $\phi$ with essentially the same form of Yukawa coupling as in the Standard Model.

Despite the similarity between (\ref{SinRSYMC}) and the lepton-Higgs sector of the Standard Model, there are also some notable differences. 
One is merely superficial: in Yukawa terms, $\phi^\dagger$ is appearing in place of $\phi$. However, this difference appears only due to our unusual assignment of antifundamental representation to a field, instead of anti-field  (made in order to make the connection with (\ref{SinSYMC}) manifest) and we might have as well relabeled $\phi \leftrightarrow \phi^\dagger$. But, more importantly, there is no independent Yukawa matrix that could differentiate between lepton generations. It cannot be directly introduced at this level, which potentially leads to a problem of mass degeneracies. Furthermore, in this simplest $U(N)$ toy model of relaxed supersymmetry, the number of fermion generations must match the rank of the unitary gauge group. Next, $U(1)$ quantum numbers are fixed by the theory, but not in the same way as in the Standard Model: in particular, $U(1)$ charge of left-handed leptons is here required to be zero, while both Higgs and the right-handed lepton have the same charge (or opposite, if we decide to relabel $\phi \leftrightarrow \phi^\dagger$). Finally, the quadratic Higgs term is not present here, reflecting that further terms must be added to fully account for the spontaneous symmetry breaking (just as in the case of standard supersymmetry). 

As our present goal is merely to illustrate the concept of relaxed supersymmetry, we will not here undertake the development of a realistic relaxed-supersymmetric analogue of the MSSM \cite{Martin:1997:MSSM} and attempt to reconcile (\ref{SinRSYMC}) with the Standard Model action. Yet, the fact that we cannot simply remove these differences by hand (while retaining relaxed supersymmetry), illustrates that mathematical rigidity remains even after discarding (\ref{anticommZ}) and (\ref{commPQ}): the action (\ref{SinRSYMC}) cannot be arbitrarily modified, since all terms are highly interrelated and constrained by the requirements of the relaxed supersymmetry. The terms are fixed by the form of the supercharge (\ref{QinRSYMC}), where the coefficients of the three terms are
over-determined: the normalization of the scalar and $\psi_R$ kinetic
terms fixes the magnitude of the second term, the cancellation of
Yukawa-type structures in the spatial momenta fixes the third completely
(including the absence of an imaginary part), and both the quartic
normalization and the cancellation of the parity-odd
$\phi^\dagger F_{ij}\phi$ terms are then satisfied without further
adjustment. In this sense, the mathematical structure turns out to be constrained in a way similar to standard supersymmetry. 

The last remark can be given a more formal treatment if we note that in the $N=1$ case (\ref{QinSYMmat}) and (\ref{QinRSYMC}) again coincide like in the Yang-Mills case, allowing us once more to see (\ref{QinRSYMC}) as a special form of generalization of abelian supersymmetry. This can be further related to the status of the constraint (\ref{anticommZ}). Namely, while the anticommutators $\{Q_R^{aA},Q_R^{bB}\}$ are not fixed by the defining relation (\ref{anticommQQPnew}), in the present models they can be evaluated explicitly. In the Yang--Mills multiplet alone
they vanish identically for all $N$; once the chiral-scalar multiplet is 	included one finds
\begin{equation}
	\{Q_R^{aA},Q_R^{bB}\}=2(P_RC^{-1})\,Z^{aA,bB}, \label{noncentralZ}
\end{equation}
with
$$Z^{aA,bB} =\frac gN \int d^3x\,
\bigl[\phi^{\dagger a}\,(\psi_R^{B})^{T} C \gamma_0\,\Psi_L^{bA}
-(aA\leftrightarrow bB)\bigr].$$

The Lorentz structure of (\ref{anticommZ}) is thus reproduced. (A priori a symmetric structure proportional to $P_R\gamma^{\mu\nu}C^{-1}$ could also appear on the right-hand side; explicit computation shows that in the present models it does not.) The operator $Z^{aA,bB}$ is, however, field-dependent and carries free gauge indices -- in particular it is neither a gauge singlet nor conserved, hence not a central charge. Nevertheless, since the anticommutator is symmetric under the simultaneous exchange of its spinor and internal indices, while $P_RC^{-1}$ is antisymmetric in the spinor indices, $Z^{aA,bB}$ is necessarily antisymmetric w.r.t.\ $aA\leftrightarrow bB$. It therefore vanishes for $N=1$, where both the gauge and the generation index take a single value. In that case, combining (\ref{noncentralZ}) with (\ref{QQPinvnew}) and with the translational invariance of the supercharges $[P_k,Q_R]=0$, algebraically (via graded Jacobi identity) implies $[H,Q_R]=0$ as well -- in other words, the model is then supersymmetric also in the standard sense.

Relaxed-supersymmetry models whose charge anticommutators reproduce the Lorentz structure of (\ref{anticommZ}), as in (\ref{noncentralZ}), with operators $Z^{aA,bB}$ that need not be central, we may call ``minimally relaxed''. For $N=1$ they become supersymmetric in the standard sense, and are thereby constrained to inherit the rigid structure of terms of their abelian supersymmetric limits -- which explains why their matter content and couplings remain as tightly restricted as in standard supersymmetry. The Yang--Mills models with and without the chiral-scalar multiplet studied here both belong to this subclass.

\section{Gauge transformations}

We shall now consider gauge transformations of relaxed-supersymmetry generators (\ref{QinRSYMC}). The action (\ref{SinRSYMC}), as well as momenta (\ref{HPinRYMmat}), are invariant with respect to standard $U(N)$ gauge transformations, that are generated by the smeared Gauss constraint:
\begin{equation}
	U[\omega]
	\equiv
	\exp\!\left[
	-2i
	\int d^3x\,
	\omega^b{}_a(\mathbf x)
	(\mathcal G_{\mathrm{YMC}}^r)^a{}_b(\mathbf x)
	\right],
\end{equation}
where $\omega^b{}_a(\mathbf x)$ is Hermitian-matrix parameter of the gauge transformation. Its action on the fields produces:
\begin{equation}
	\begin{aligned}
		A_i'
		&=
		U[\omega]A_i U[\omega]^{-1}
		=
		V A_i V^{-1}
		-
		\frac{i}{g}
		(\partial_i V) V^{-1},
		\\
		\phi'_a
		&=
		U[\omega]\phi_a U[\omega]^{-1}
		=
		\phi_b(V^{-1})^b{}_a,
		\\
		(\Psi_L')^{aA}
		&=
		U[\omega](\Psi_L)^{aA} U[\omega]^{-1}
		=
		\widehat V^a{}_b (\Psi_L)^{bA},
		\\
		(\psi_R')^A
		&=
		U[\omega](\psi_R)^A U[\omega]^{-1}
		=
		e^{-i g \operatorname{Tr}(\omega)/N} (\psi_R)^A ,
	\end{aligned}
\end{equation}
where $V(\mathbf x) \equiv e^{\,ig\omega(\mathbf x)}$ is a unitary matrix and $\widehat V(\mathbf x)$ is its $SU(N)$ part obtained by exponentiating the traceless component $\widehat \omega^b{}_a(\mathbf x) = \omega^b{}_a(\mathbf x) - \frac 1N \delta ^b_a \omega^c{}_c(\mathbf x)$. 

If we denote density of the supercharge as $\suc_0^{aA}$, so that (\ref{QinRSYMC}) can be shortened as $Q_R^{aA} = \int d^3x\, \suc_0^{aA}$, under gauge transformations it will transform as
\begin{equation}
	\suc_{0}^{\prime aA}
	=
	U[\omega]\suc_0^{aA} U[\omega]^{-1}
	=
	\widehat V^a{}_b \suc_0^{bA}.
\end{equation}
Here $\widehat V$ appears instead of $V$ because all terms in (\ref{QinRSYMC}) have vanishing overall $U(1)$ charge. Consequently, operators $Q_R^{aA}$ substantially change their form under gauge transformations:
\begin{equation}
	Q_R^{\prime aA} =
	U[\omega] Q_R^{aA} U[\omega]^{-1}
	= \int d^3x\,
	\widehat V^a{}_b(\mathbf x) \suc_0^{bA}(\mathbf x),
\end{equation}
revealing that setting the specific form (\ref{QinRSYMC}) requires gauge fixing. Thus, it is more elegant to define  generators of the relaxed supersymmetry in the following way:
\begin{equation}
 	Q_R^{\bar aA}[\SUtet] = \int d^3x\, \SUtet^{\bar a}{}_a(\mathbf x) \suc_0^{aA}(\mathbf x). \label{QgaugeCov}
\end{equation}
Here, $\SUtet^{\bar a}{}_a(\mathbf x)$ is a coordinate-dependent $SU(N)$ matrix, which represents an internal $SU(N)$ frame, loosely analogous to a tetrad. The overlined index $\bar a$ is a ``global'' $SU(N)$ index and is invariant with respect to (local) gauge transformations. With this definition, the class of operators (\ref{QgaugeCov}) now retains the overall form under gauge transformations:  
\begin{equation}
	Q_R^{\prime \bar aA}[\SUtet] =
	U[\omega] Q_R^{\bar aA} U[\omega]^{-1}
	= Q_R^{\bar aA}[\SUtet \widehat V].
\end{equation}
The defining relation (\ref{anticommQQPnew}) or, more precisely, its generalization (\ref{QQPinvnew}) still holds, provided that the summation is now carried over the ``global'' index:
\begin{equation}
	\left\{ Q^{\bar aA}_R[\SUtet], 
	\overline{Q_R}_{\bar aA}[\SUtet] \right\}
	=2\bigl(P_R \gamma^\mu \bigr) P_\mu. \label{anticommQQPnewGlobal}
\end{equation}
This follows directly if one gauge-transforms (\ref{QQPinvnew}) and takes into account that the right-hand side is gauge invariant. Of course, it can be also confirmed by explicit calculation, when the extra terms containing derivatives of $\SUtet$ vanish due to identity $\operatorname{tr}(\SUtet^{-1} \widehat D_k \SUtet) = 0$, valid due to $\det \SUtet =1$ and $\operatorname{tr}\widehat A_k=0$.  

While in (\ref{QgaugeCov}) we introduced $\SUtet$ simply as an $SU(N)$-valued parameter function, it is an interesting question whether it can be promoted to a full-fledged quantum field, that would transform as $\SUtet \rightarrow \SUtet \widehat V^{-1}$ and make operators $Q^{\bar aA}_R$ gauge invariant under local transformations. However, we will not pursue this idea in the present paper.

\section{Summary}

We presented a different interpretation of supersymmetry, directly relating it to internal gauge symmetries. In this view, supersymmetry generators are the fundamental entities, labeled not only by a spinor index but also by an index of a representation of an internal gauge group and by a generation index. Spacetime momenta can here be seen as derived entities, formed as gauge-invariant bosonic combinations of supersymmetry generators -- as given by the relation (\ref{QQPinvnew}), superseding (\ref{anticommQQPnew}).
Supersymmetry generators are translation-invariant (in the fixed gauge frame), but are not required to commute with Hamiltonian (nor to generate symmetries of the action). Relaxation of this constraint opens the possibility of combining existing known particles into supermultiplets, as was illustrated in the paper.

Both models discussed here belong to the ``minimally-relaxed'' subclass of all relaxed supersymmetric models that satisfy (\ref{noncentralZ}) with possibly non-central $Z^{aA,bB}$. For this class, the relaxed supersymmetry can be seen as an entirely different, third way to generalize expressions of the standard non-extended abelian Yang-Mills supersymmetry with matter. Two paths of generalization are commonly studied \cite{Strathdee:1987:ClassificationNSUSY}: i) enlarging only the gauge group from $U(1)$ to $U(N)$, while keeping the number of supersymmetry generators and thus introducing gauginos and sfermions with properties that do not match any known particles; ii) extending the number of supersymmetries by adding a $U(N)$ index on generators, but without any direct relation with the internal gauge group. Both approaches are at odds with experimental results, due to prediction of a slew of hitherto unobserved particles. The third way that we described here extends the abelian gauge group by assigning an $SU(N)$ gauge-group index directly to the generators of the supersymmetry, after which only gauge invariant combinations of charge anticommutators can remain symmetries of the action.  

The obtained relaxed-supersymmetric model of leptons interacting with scalar and $U(N)$ Yang-Mills fields has a strong resemblance in field content and interaction pattern with lepton-Higgs sector of the Standard Model. It also has some important differences that cannot be trivially removed by hand -- also a reminder that the approach is mathematically strongly constrained: its relative coefficients and field multiplicities are fixed by the defining relation. 

However, the relaxed supersymmetry offers more freedom in constructing the theory than the standard supersymmetry. The relaxed setting may also admit further couplings -- for instance additionally (or alternatively) relating $\Psi_L^{aA}$ to scalars in a two-index representation -- which could in principle allow a non-zero trace-$U(1)$ charge for $\Psi_L^{aA}$ and, in a more complicated setting, lift the degeneracy of the Yukawa couplings (such constructions are generally no longer minimally-relaxed). We emphasize that this has not been carried out here: any such extension potentially alters the counting that fixes the number of generations, and the defining relation would have to be re-established.

We close by stating explicitly some important issues that remain to be investigated, but do not fall into the scope of this paper. The individual operators $Q^{aA}_\alpha$ are not conserved and do not generate symmetries of the action; only the fully-contracted combinations (\ref{QQPinvnew}) do. Accordingly, no corresponding Ward identities are available here, and whether the rigid relations among the couplings that (\ref{QQPinvnew}) imposes survive radiative corrections remains to be investigated (the obtained relations have not been examined beyond the classical level). Furthermore, the anticommutators not fixed by (\ref{QQPinvnew}) have not been systematically analyzed, so no claim is made about closure of the full algebraic structure. Which of the the mathematical advantages offered by the standard supersymmetry  are retained in the relaxed supersymmetry context, and what are the renormalization properties of relaxed models, also needs to be explored. It is only certain that, due to (\ref{QQPinvnew}), the positivity of energy remains also a feature of relaxed supersymmetry. 

On the phenomenological side, the simplest $U(N)$ realization ties the number of fermion generations to $N$, assigns $U(1)$ charges differently from the Standard Model, leaves the lepton generations mass-degenerate, and lacks the terms required for spontaneous symmetry breaking; anomaly cancellation has not been addressed. The models presented here are therefore intended as a proof of concept and an illustration of the idea of relaxed supersymmetry, rather than as realistic candidates. While the development of a realistic model is well beyond present scope, we only note that a natural direction would be towards some of the GUT models \cite{Langacker:1980:GUT}, with a more unified gauge-group structure.

\bibliography{references}

\end{document}